\documentclass[twocolumn]{aastex61}

\received{January 1, 2018}
\revised{January 7, 2018}
\accepted{\today}

\submitjournal{ApJ Letters}

\shorttitle{Gas accretion within the dust cavity in AB Aur}
\shortauthors{Rivi\`ere-Marichalar et al.}

\begin{document}

\title{Gas accretion within the dust cavity in AB Aur\footnote{Based on observations carried out with the IRAM  Northern
Extended Millimeter Array (NOEMA). IRAM is supported by INSU/CNRS (France), MPG (Germany), and IGN (Spain).}}

\correspondingauthor{Pablo Rivi\`ere-Marichalar}
\email{p.riviere@oan.es}

\author{Pablo Rivi\`ere-Marichalar}
\affil{Observatorio Astron\'omico Nacional (OAN, IGN), Calle Alfonso XII 3\, E-28014 Madrid, Spain}

\author{Asunci\'on Fuente}
\affiliation{Observatorio Astron\'omico Nacional (OAN, IGN), Calle Alfonso XII 3\, E-28014 Madrid, Spain}

\author{Cl\'ement Baruteau}
\affiliation{CNRS / Institut de Recherche en Astrophysique et Plan\'etologie, 14 avenue Edouard Belin, F-31400 Toulouse, France}

\author{Roberto Neri}
\affiliation{Institut de Radioastronomie Millim\'etrique, 300 rue de la Piscine, F-38406 Saint-Martin d'H\`eres, France}

\author{Sandra P. Trevi\~no-Morales}
\affiliation{Dept.\ of Space, Earth and Environment, Chalmers University of Technology, Onsala Space Observatory, SE-439 92 Onsala, Sweden}

\author{Andr\'es Carmona}
\affiliation{CNRS / Institut de Recherche en Astrophysique et Plan\'etologie, 14 avenue Edouard Belin, F-31400 Toulouse, France}

\author{Marcelino Ag\'undez}
\affiliation{Instituto de F\'isica Fundamental, CSIC, Serrano 123, E-28006  Madrid, Spain}

\author{Rafael Bachiller}
\affiliation{Observatorio Astron\'omico Nacional (OAN, IGN), Calle Alfonso XII 3\, E-28014 Madrid, Spain}

\begin{abstract}

AB Aur is a Herbig Ae star hosting a well-known transitional disk. Because of its proximity and low inclination angle, it is an excellent object to study planet formation. Our goal is to investigate the chemistry and dynamics of the molecular gas component in the AB Aur disk, and its relation with the prominent horseshoe shape observed in continuum mm emission. We used the NOEMA interferometer to map with high angular resolution the J = 3-2 lines of HCO$\rm ^+$ and HCN. By combining both, we can gain insight into the AB Aur disk structure. Chemical segregation is observed in the AB Aur disk: HCO$^+$ shows intense emission toward the star position, at least one bright molecular bridge within the dust cavity, and ring-like emission at larger radii, while HCN is only detected in an annular ring that is coincident with the dust ring and presents an intense peak close to the dust trap. We use HCO$^+$ to investigate the gas dynamics inside the cavity. The observed bright HCO$^+$ bridge connects the compact central source with the outer dusty ring. This bridge can be interpreted as an accretion flow from the outer ring to the inner disk/jet system proving gas accretion through the cavity.

\end{abstract}


\section{Introduction} \label{sec:intro}

AB Aur is a Herbig A0-A1 star \citep{Hernandez2004} located at 162.9 $\rm \pm$ 1.5 pc from the Sun \citep{Gaia2018}, which hosts a transitional disk. Near-IR observations with HST-STIS and SUBARU-CIAO showed that the disk presents spiral-arm signatures \citep{Grady1999, Fukagawa2004, Hashimoto2011}, which could be due to  one or several giant-forming planets. Observations of CO J = 2-1, $^{13}$CO J = 2-1, and continuum at 1.3 mm with PdBI showed the presence of an inner dust cavity extending out to 70-100 au, surrounded by an asymmetric dusty ring \citep{Pietu2005,Tang2012, Fuente2017}.  Higher spatial resolution sub-mm imaging resolved the inner disk \citep{Tang2012}. Furthermore, observations with the eVLA revealed the presence of a radio jet \citep{Rodriguez2014}, which is consistent with the high levels of accretion derived for the system \citep{GarciaLopez2006,Salyk2013}. High angular resolution (0$\arcsec$.11$\times$0$\arcsec$.08) ALMA images of the CO J = 2-1 line in AB Aur showed the existence of two prominent CO spiral arms at a radius of 0$\arcsec$.3 \citep{Tang2017}. Whether the spiral arms are connected with the outer dust ring remains to be clarified, as the aforementioned images were subject to strong filtering due to the limited \textit{uv}-plane coverage, precluding the detection of large structures.

Our team is carrying out a long-term study of this prototypical disk. On the basis of the comparison of the continuum NOEMA maps at 2.2 and 1.1 mm with two-fluid (gas+dust) hydrodynamical simulations \citep{Baruteau2016,Zhu2016}, \citet{Fuente2017} proposed the existence of a decaying vortex toward the mm dust emission peak, which is needed to explain, in the presence of azimuthal dust trapping, why the azimuthal contrast ratio along the (sub)mm ring is smaller at 2.2 mm than at 1.1 mm. The vortex is assumed to have formed at the outer edge of the gap carved by a massive planet, and its decay is caused by the disk turbulent viscosity. The study constrained the total mass of solid particles in the ring, which was found to be of the order of 30 $ M_{\earth}$. 

A handful of species were detected toward  AB Aur using single-dish and interferometric  observations \citep{Fuente2010,Pacheco2015,Pacheco2016}. Interferometric images of $^{13}$CO J = 2-1, C$^{18}$O J = 2-1, SO J = 5$_6$-4$_5$, and H$_2$CO J = 3$_{0,3}$-2$_{02}$ with an angular resolution of $\sim$1$\arcsec$.6$\times$1$\arcsec$.5 testify to the complex gas chemistry occurring in the ring, which was found to be related with the gas dynamics. In particular, \citet{Pacheco2016} found a local minimum in the SO abundance toward the dust trap, which was interpreted as the gas vortex chemical footprint.

In this Letter we extend our chemical and dynamical study to smaller spatial scales by using higher resolution images of the HCO$^+$ J = 3-2 and HCN J = 3-2 lines, which provide new insights into the evolution of the gas in this protoplanetary disk.
 
\section{Observations and data reduction}

We present high spatial resolution $\sim$0$\arcsec$.4 ($\sim$60 AU at the distance of AB Aur) observations of the HCO$^+$ J = 3-2 and HCN J = 3-2 lines in the AB Aur disk. The observations were carried out with the NOEMA interferometer in their A and C configurations. The observations in C configuration were carried out in 2017 January 1-3. During these observations, the wideband correlator WIDEX ($\Delta \nu$=2 MHz) was used to cover the $\sim$2 GHz receiver band, and two correlator units providing a higher spectral resolution of $\sim$39 kHz were located at  the frequencies of the HCO$^+$ J = 3-2 and HCN J = 3-2 lines, respectively. The region empty of emission channels was used to build the 1.12 mm continuum published by \citet{Fuente2017}. The observations in A configuration were done in 2018 March, using the new correlator PolyFix, which provides a total band of $\sim$ 7.74 GHz with a spectral resolution of $\Delta \nu$=2 MHz. In addition, higher spectral  resolution $\Delta \nu$=65 kHz chunks were located at the frequencies of the HCO$^+$ and HCN lines. All data from A and C configurations were resampled to a frequency resolution of 89 kHz and merged to build the A+C table of visibilities used to create the images shown in this Letter.  In order to optimize the spatial resolution, we have applied uniform weighting to produce the HCO$^+$ J = 3-2 image. In the case of the HCN J = 3-2 line, we prefer to use natural weighting because the weaker line emission. Self-calibration was applied to improve the sensitivity and dynamical range. Data calibration and imaging were done using the package \texttt{GILDAS}\footnote{See \texttt{http://www.iram.fr/IRAMFR/GILDAS} for  more information about the GILDAS  softwares.}\texttt{/CLASS} software.

\begin{figure*}[!t]
\begin{center}
 \includegraphics[scale=0.24]{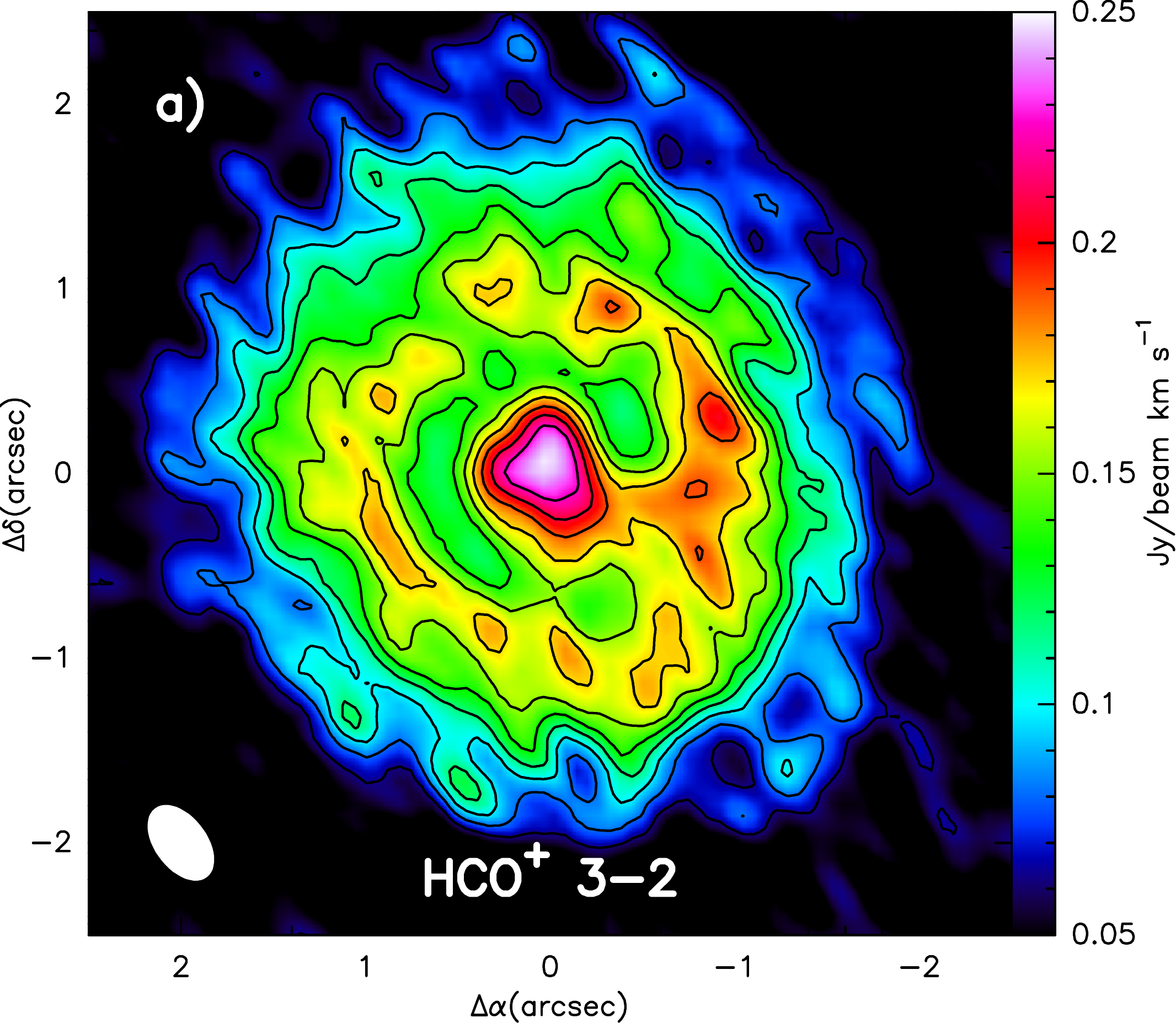}
 \includegraphics[scale=0.24]{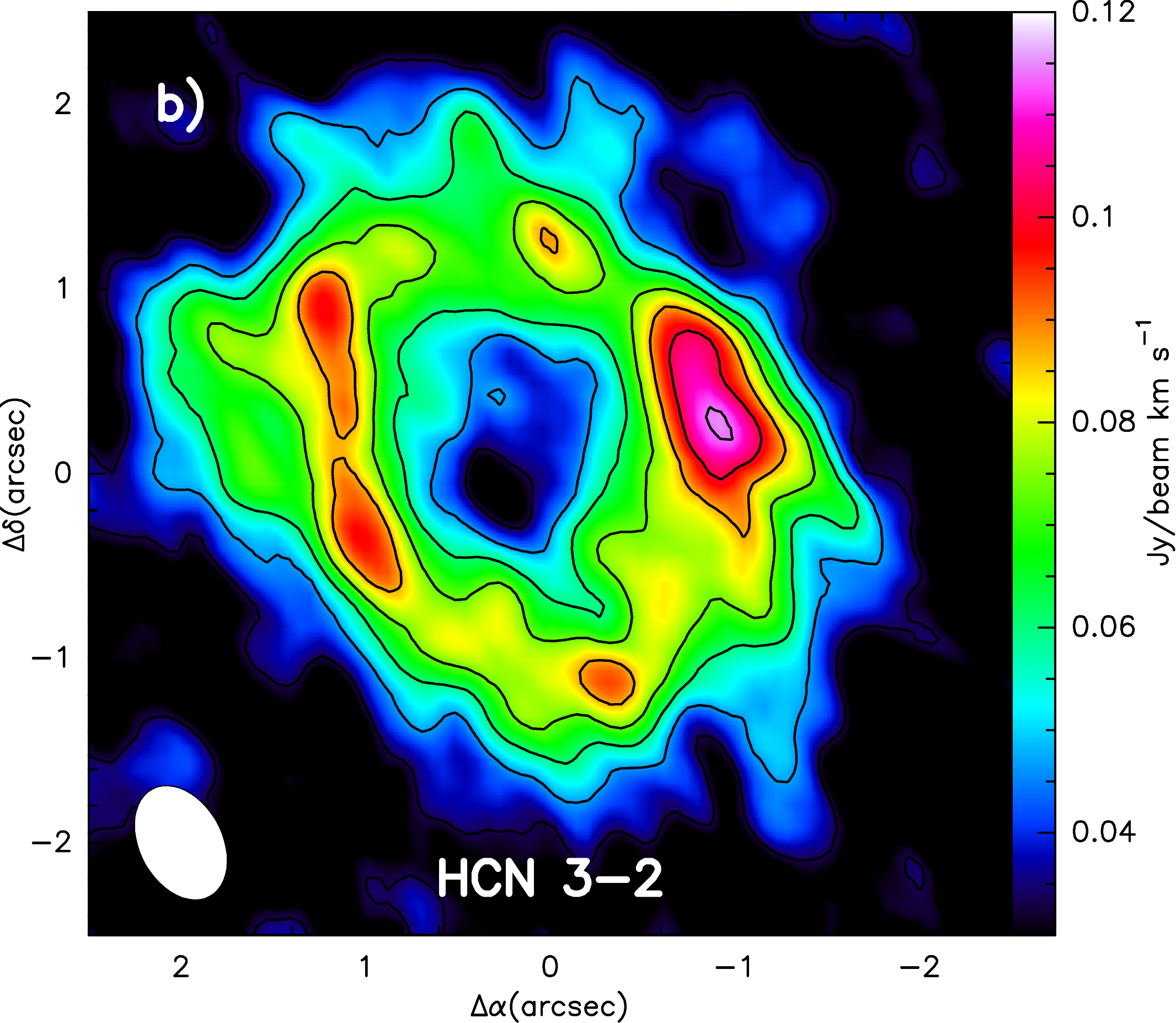}
 \includegraphics[scale=0.24]{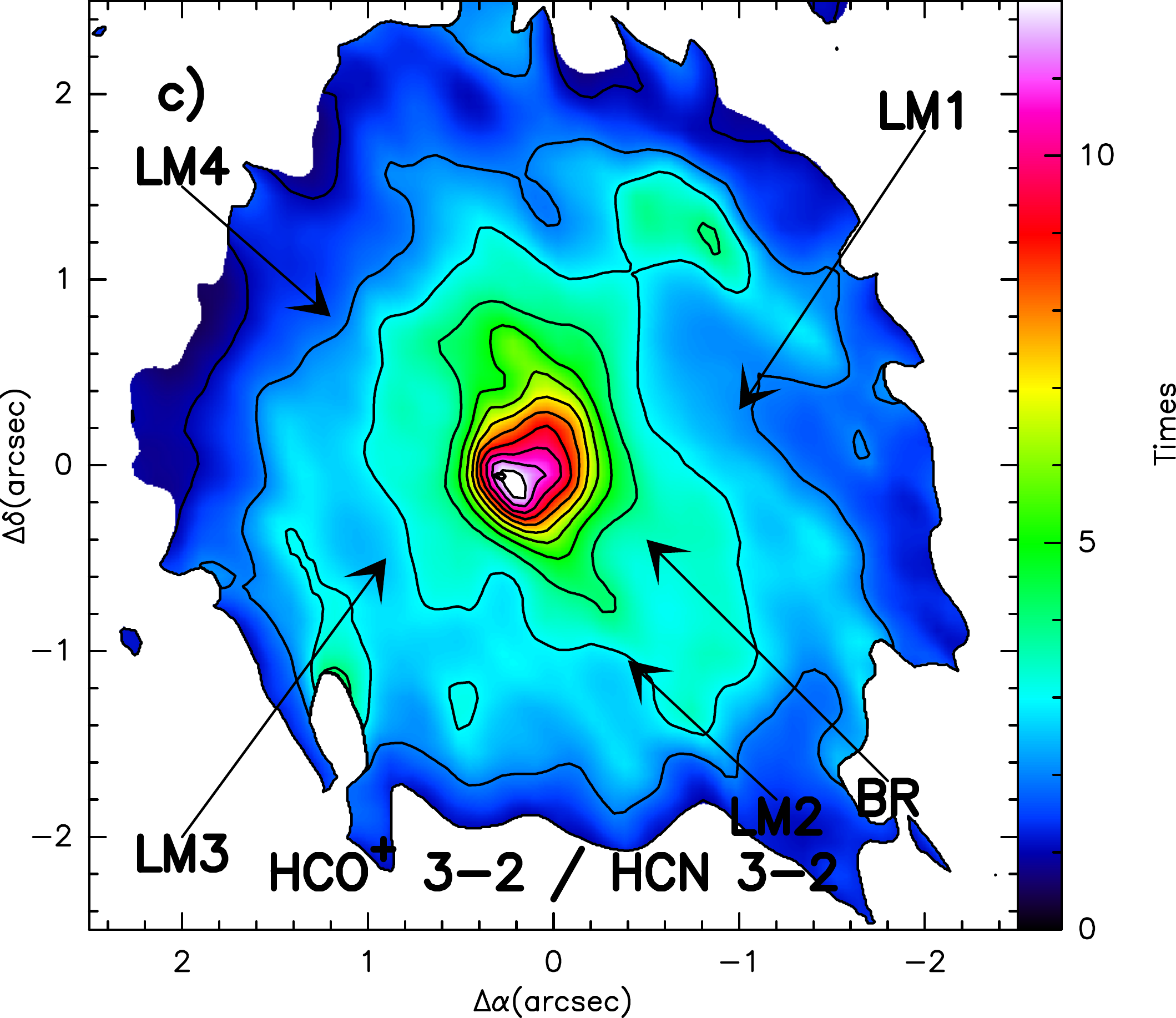}\\
 \includegraphics[scale=0.24]{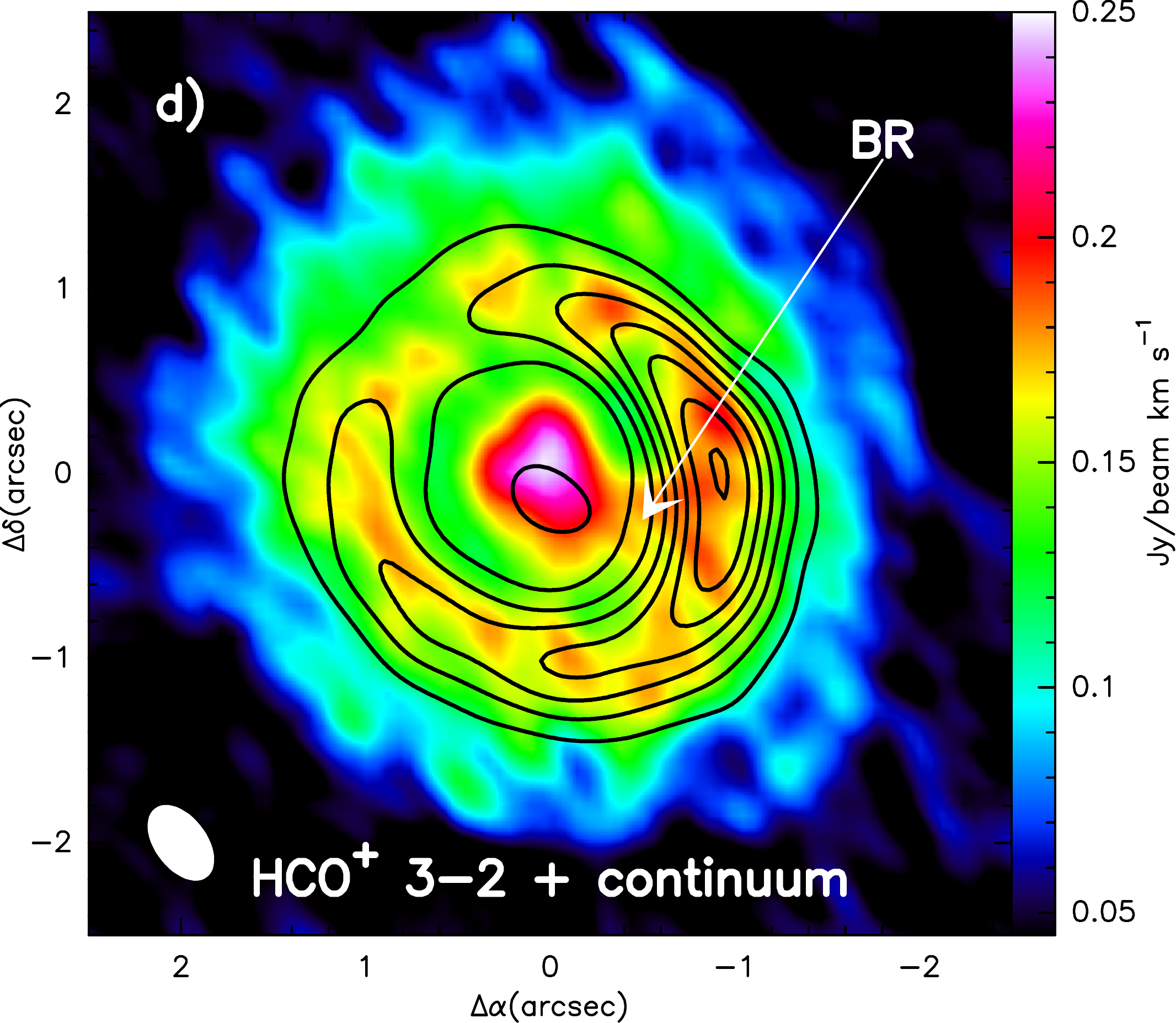}
 \includegraphics[scale=0.24]{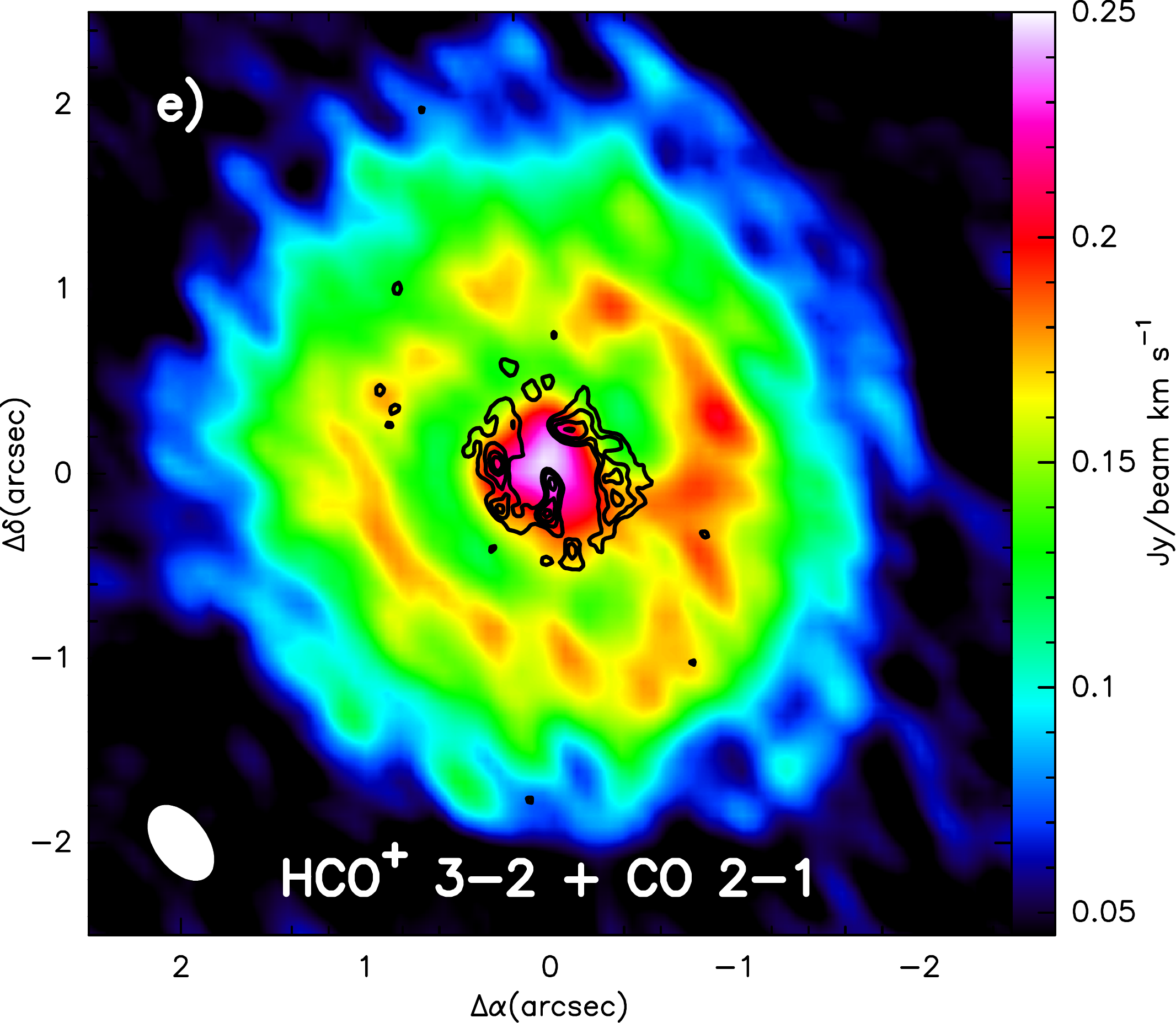}
 \includegraphics[scale=0.24]{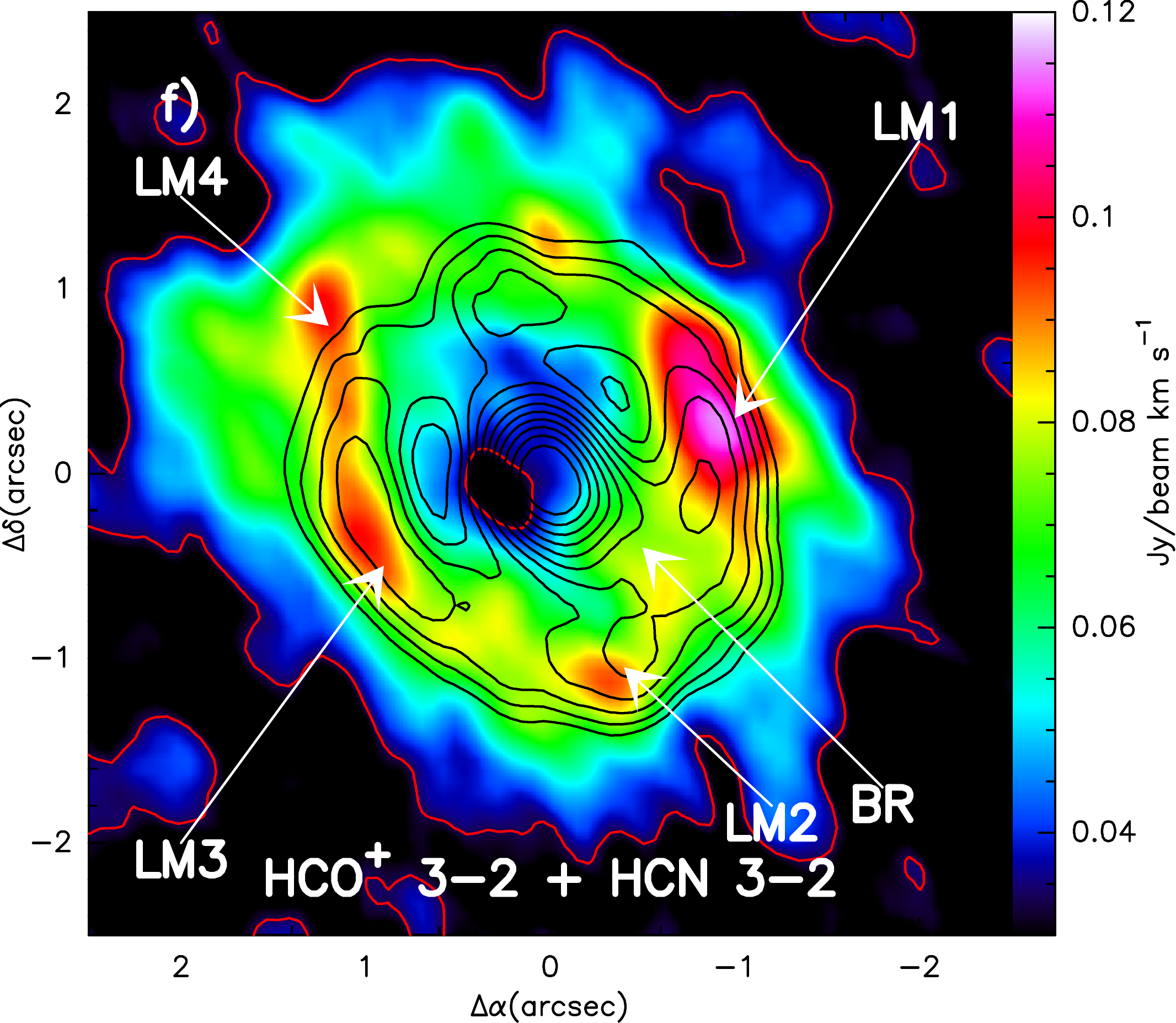}\\
 \caption{(a) Integrated intensity map of HCO$\rm ^+$ using uniform weight. Contour levels are 5$\sigma$ to 40$\sigma$ in steps of 4$\sigma$. (b) Integrated intensity map of HCN using natural weight. Contour levels are 5$\sigma$ to 20$\sigma$ in steps of 1$\sigma$. (c) Map of HCO$\rm ^+$ over HCN, using only channels with S/N$\rm > 5$. Contour levels are 2 to 24 in steps of 4. (d) Integrated intensity map of HCO$\rm ^+$ with continuum intensity map at 1.12 mm overlaid as white contours. Continuum contour levels are 15$\sigma$ to 75$\sigma$ in steps of 10$\sigma$. (e) CO J = 3-2 emission contours from Tang et al. (2017) overlayed on the HCO+ J = 3-2 image. (f) HCN integrated intensity map with HCO$\rm ^+$ overlaid as black contours (natural weights). HCO$\rm ^+$ contour levels are 20$\sigma$ to 40$\sigma$ in steps of 1$\sigma$. The 5$\sigma$ level is also shown. Interesting positions in the maps are labeled in white. Integrated line intensity maps were built by integrating velocity channels between +2 and +10 km $\rm s^{-1}$. Beam sizes are shown in the bottom-left corner of each plot. }
\label{Fig:IntensityMaps}
\end{center}
\end{figure*}

\section{Results}
As Figure \ref{Fig:IntensityMaps} shows, the integrated intensity maps of the HCO$\rm ^+$ J = 3-2 and HCN J = 3-2 lines unveil important differences in the distribution of the HCO$\rm ^+$ and HCN emissions in the radial and azimuthal directions. While HCN is only detected (above 5$\sigma$) in an annular ring that is spatially coincident with the dust ring (hereafter, outer ring), HCO$^+$ is detected in the outer ring and also features intense and compact emission toward the star position. Additionally, a bright HCO$^+$ bridge connecting the compact central source with the outer ring (labeled BR in Figure \ref{Fig:IntensityMaps}) is observed. To test whether the bridge is due to an elongation of the central emission along the beam major axis, we show in Figure \ref{Fig:SpectralComparison} a comparison of the spatially integrated line profiles of HCO$^+$ at different positions. The line profile along the bridge (Figure \ref{Fig:SpectralComparison}(b)) is quite different from that of the central emission (Figure \ref{Fig:SpectralComparison}(a)): while the profile in the innermost region peaks at ~6.7 km s$\rm ^{-1}$ (close to the system velocity), with an FWHM of ~1.4 km s$\rm ^{-1}$, the line profile at the bridge location is two-peaked, with a red-shifted wing at ~7.3 km s$\rm ^{-1}$ with an FWHM ~0.8 km s$\rm ^{-1}$ and a blueshifted broad and fainter wing at $\sim$ 5.3 km s$\rm ^{-1}$. The difference in the line profiles shows that the emission seen at the location of the bridge is not the elongation by the beam of the emission from the inner disk, thus a feature on its own. 

In Figure \ref{Fig:IntensityMaps}e we compare our HCO$\rm ^+$ J = 3-2 integrated intensity map with the CO J = 2-1 observations by \cite{Tang2017}. The southwestern spiral-arm observed in CO J = 2-1 overlaps with the position of the HCO$\rm ^+$ J = 3-2 bridge in its southern side, indicating a possible connection between both structures.

We also find significant differences between the HCO$^+$ and HCN emission maps along the azimuthal direction. In 
order to illustrate these differences we have labeled the HCN clumps along the outer ring as LM1 to LM4 (see Figure \ref{Fig:IntensityMaps}). The HCO$^+$ emission in the outer ring follows quite well the dust emission  with intensity variations $<$ 2$\sigma$  along it. This is in agreement with what we expect from two-fluid simulations \citep{Fuente2017} in which the gas surface density is quite uniform along the ring, while large particles are still accumulated in the dust trap. We note that we measure an intensity contrast of a factor of $\sim$3 in the dust 1.12mm continuum emission. In contrast, the HCN emission seems to follow the spatial distribution of large particles,  with the most intense clump, LM1, being spatially coincident with the dust trap. In the eastern-half disk where the 1.12mm continuum emission presents its minimum, the HCN emission seems to come from the external parts of the ring (see LM4 in Figure \ref{Fig:IntensityMaps}). Given the limited sensitivity of our maps it is important to note that LM2, LM3, and LM4 could be due to strong noise fluctuations. Although the HCO$^+$ and HCN are both high-density tracers, the two molecules present a different chemistry and are expected to come from different layers within the disk \citep{Pacheco2015}. HCN is known to be abundant in dense cold regions where CO and HCO+ are frozen on grains \citep{Fuente2019}. In addition, the emission of HCO$^+$ might be optically thick toward the outer ring.

\section{Discussion}
\subsection{Gas inside the mm dust cavity}
Low-resolution spectral observations of AB Aur are available from about 1 $\mu$m up to about 400 $\mu$m \citep[][]{Woitke2018}. The detection of several O I, C II, C, OH, and ro-vibrational lines of CO and H$_2$ prove the existence of warm gas \citep{Brittain2003, Bitner2007, Riviere2016}. The origin (inner disk, outflow, gas filling the cavity)  of these line emissions is uncertain because of the moderate angular and spectral resolutions of the observations. Based on observations of the [O I] 63 $\mu$m line, \citet{Riviere2016}  concluded that part of the atomic oxygen emission is coming from the jet.

\begin{figure}[!h]
\begin{center}
 \includegraphics[scale=0.6, trim=1mm 8mm 0mm 10mm, clip]{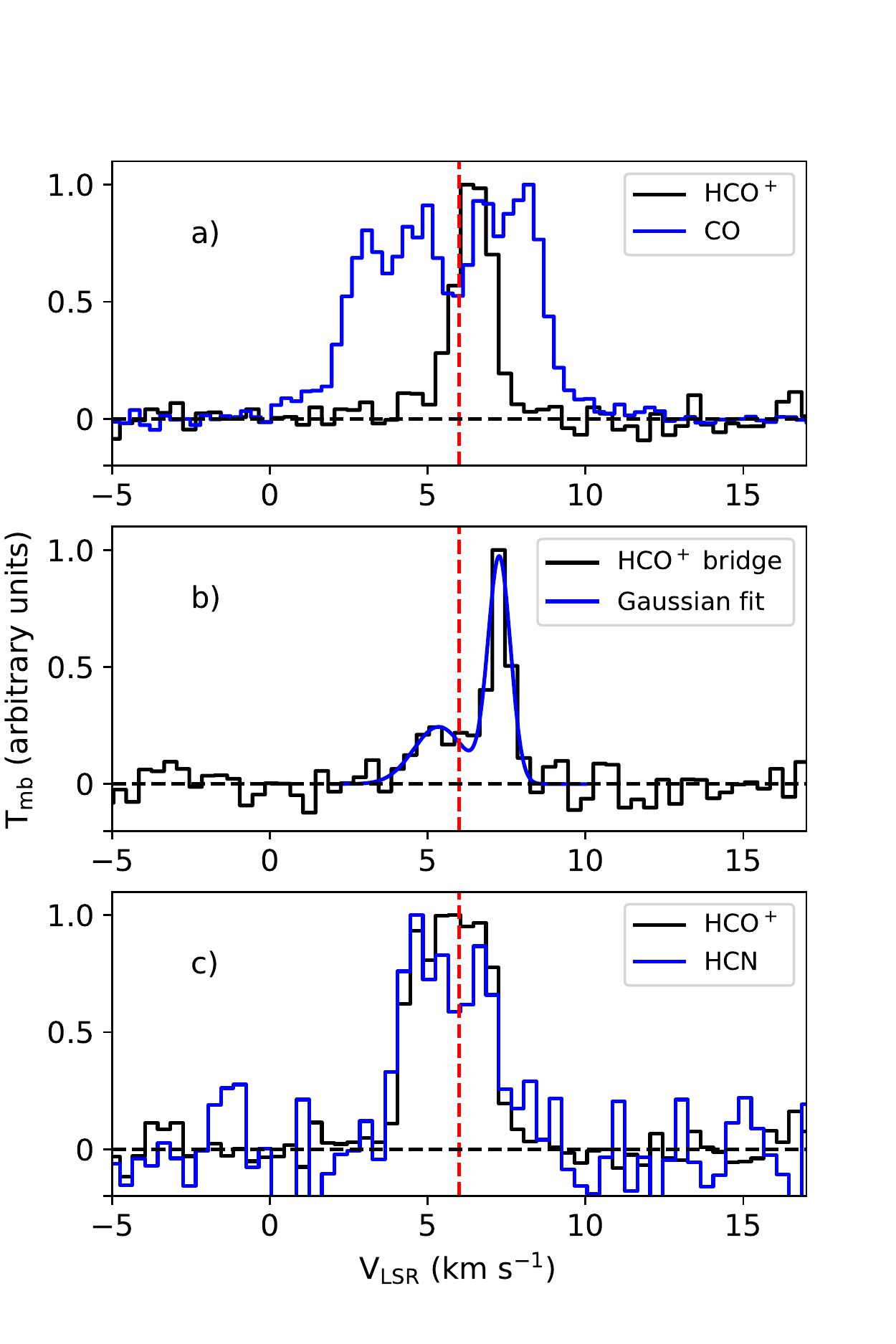}\\
  \includegraphics[scale=0.3, trim=0mm 0mm 0mm 0mm, clip]{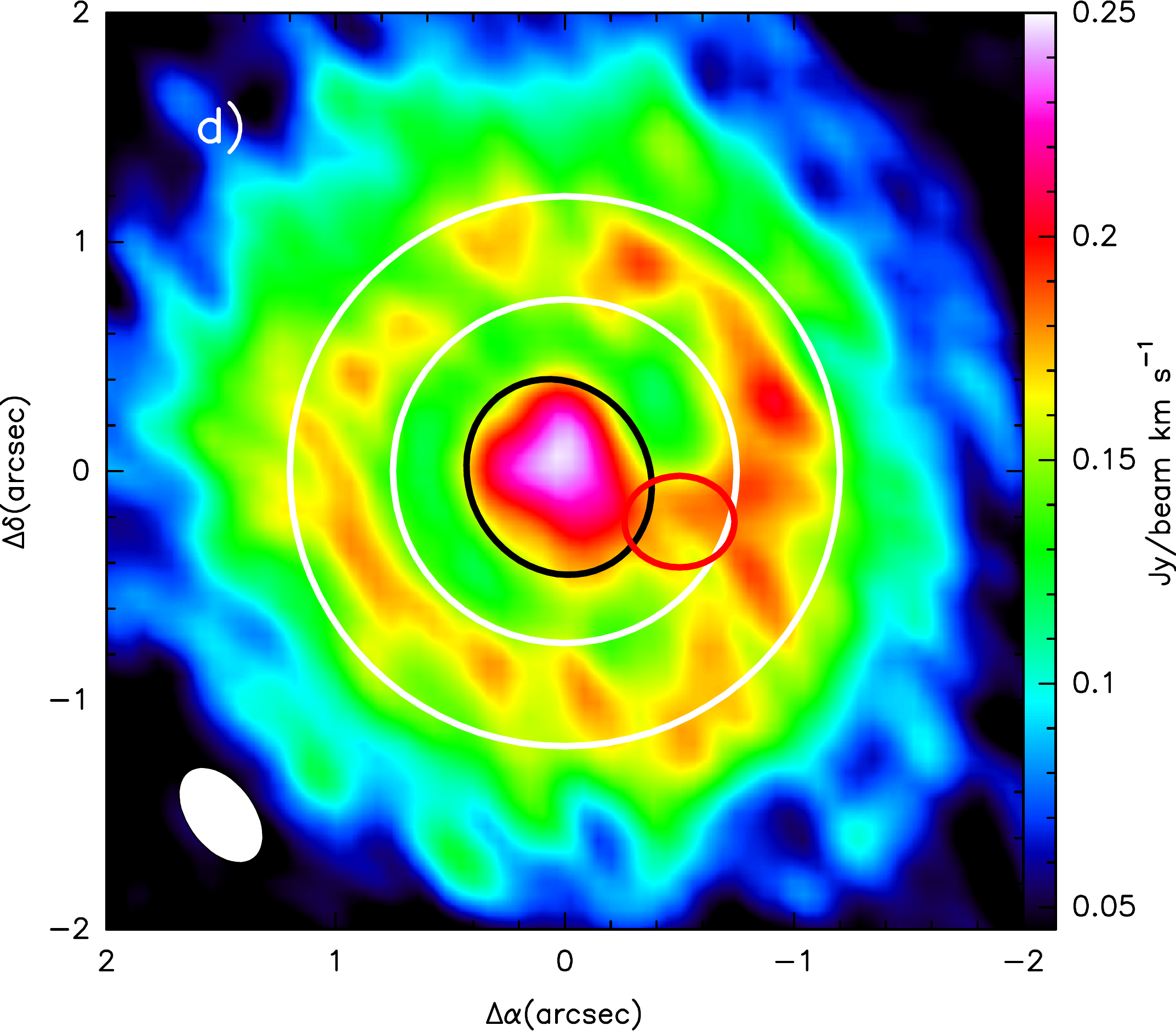}
 \caption{(a) Comparison of the spatially integrated line profile of CO J = 2-1 (blue) from \cite{Tang2017} and HCO$\rm ^+$ J = 3-2 (black) over the inner regions. The vertical dashed line displays the system velocity. (b) Spatially integrated HCO$\rm ^+$ line profile over the bridge position. (c) Comparison of spatially integrated HCN and HCO$\rm ^+$ line profiles over the outer ring. All the spectra have been scaled to their maximum values to allow for an easy comparison. (d) Location of the regions used to obtain the integrated spectra are shown as a black ellipse (innermost regions), a red ellipse (bridge), and white circles (ring). }
 \label{Fig:SpectralComparison}
\end{center}
\end{figure}

Interferometric (sub-)millimeter images provide insights into the gas and dust distributions. Out of all the molecular lines imaged with millimeter interferometers in AB Auriga \citep{Pietu2005,Tang2012,Pacheco2016,Tang2017}, only the high-resolution CO J = 2-1 image reported by \citet{Tang2017} and our HCO$^+$ J = 3-2 image show evidence of molecular 
emission within the cavity. These two lines present, however, different line profiles. Figure \ref{Fig:SpectralComparison}(a) shows integrated CO J = 2-1  and HCO$^+$ J = 3-2 spectra in the innermost region (see Figure \ref{Fig:SpectralComparison}(d). The line profile of the HCO$^+$ line shows a narrow, $\Delta v$ = 1.4~km~s$^{-1}$ component at 6.7 km s$^{-1}$ 
on top of broad wings covering the $\sim$0$-$10 km s$^{-1}$ velocity range. Although the CO J = 2-1 line covers the same velocity range, its profile is flat-topped. As commented by \citet{Tang2017}, the flat CO line profile can be
due to self-absorption and spatial filtering effects. It should be noted that recent observations of CO J = 3-2 and HCO$^+$ J = 3-2 in AA Tau by \citet{Loomis2017} showed different spatial distribution of  the HCO$^+$ J = 3-2, and  CO J = 3-2 emissions, suggesting that they are indeed coming from different regions.

The poor knowledge of the physical conditions of the gas within the cavity ($r <$ 80 au) makes it difficult to
derive molecular abundances. \citet{Woitke2018} calculated a gas mass of 4.2$\times$10$^{-4}$ $M_{\odot}$ for the inner disk, with a column density of $N_H$$\sim 3\times$10$^{22}$ cm$^{-2}$ at  $r \sim$ 40 au. 
Assuming optically thin emission and that the HCO$^+$ J = 3-2 line is thermalized at $T_k$ = 150 K \citep{Tang2017}, we derive $N$(HCO$^+$)=3$\times$10$^{13}$~cm$^{-2}$, i.e., an HCO$^+$ abundance of $\sim$ 10$^{-9}$ within the cavity,  which is compatible with typical values in photo-dissociation regions \citep[see, e.g.,][]{Fuente2003,Goicoechea2017, Ginard2012}. The HCO$^+$ abundance might be larger if the HCO$^+$ J = 3-2 emission is optically thick.

\subsection{HCO$^+$/HCN: A chemical tool}
Dramatic chemical segregation is found in the radial direction and along azimuth within the  AB Aur disk. The absence of large grains within the cavity allows the UV irradiation from the star to penetrate through it and photo-dissociate molecules. Very few species are expected to survive in highly irradiated gas. H$_2$ and CO are abundant enough to self-shield the stellar radiation \citep[see, e.g.,][]{Sternberg1995}. In addition, small hydrides (CH, CH$^+$, NH), reactive ions (CO$^+$, HOC$^+$), and simple oxygenated species such as  OH and HCO$^+$, are detectable in the highly irradiated layers of photon-dominated regions \citep{Fuente2003, Goicoechea2017}. Because it is easily observable at millimeter wavelengths,  HCO$^+$ is an excellent candidate to investigate the gas dynamics within the cavities of nearby transition disks.

Our observations suggest that the HCO$^+$/HCN ratio is a useful chemical tool in determining the structure of transitional disks.  In the case of unresolved disks, a high HCO$^+$/HCN intensity ratio might be  associated with the existence of inner dust cavities full of dense gas. We note that, together with AB Aur,  HD 142527 and AA Tau were the disks with a higher HCO$^+$ J = 3-2 / HCN J = 3-2 ratio in the compilation of \cite{Pacheco2015}. Both systems show filamentary gas emission through their dust cavities \citep{Cassasus2013, Loomis2017}. The other disk with a high ratio, GM Aur, has not been studied in detail. This result is also in line with recent results by \citet{Alonso2018} toward the Herbig Be star  R~Mon.

\subsection{Gas kinematics}\label{Sec:Gas_kin}
\begin{figure*}[]
\centering
 \includegraphics[scale=0.42, trim=5mm 0mm 15mm 0mm, clip]{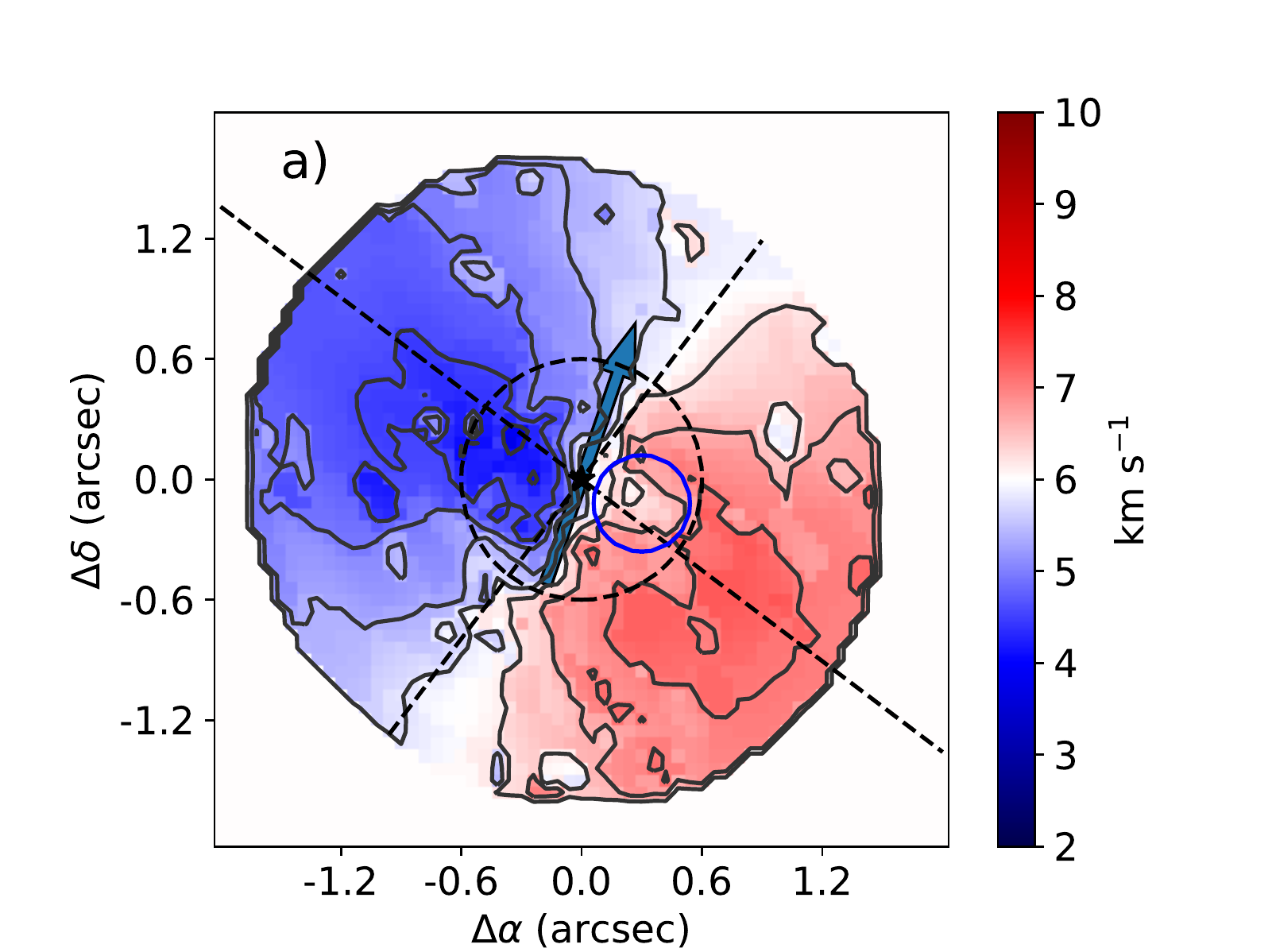}\includegraphics[scale=0.42, trim=5mm 0mm 15mm 0mm, clip]{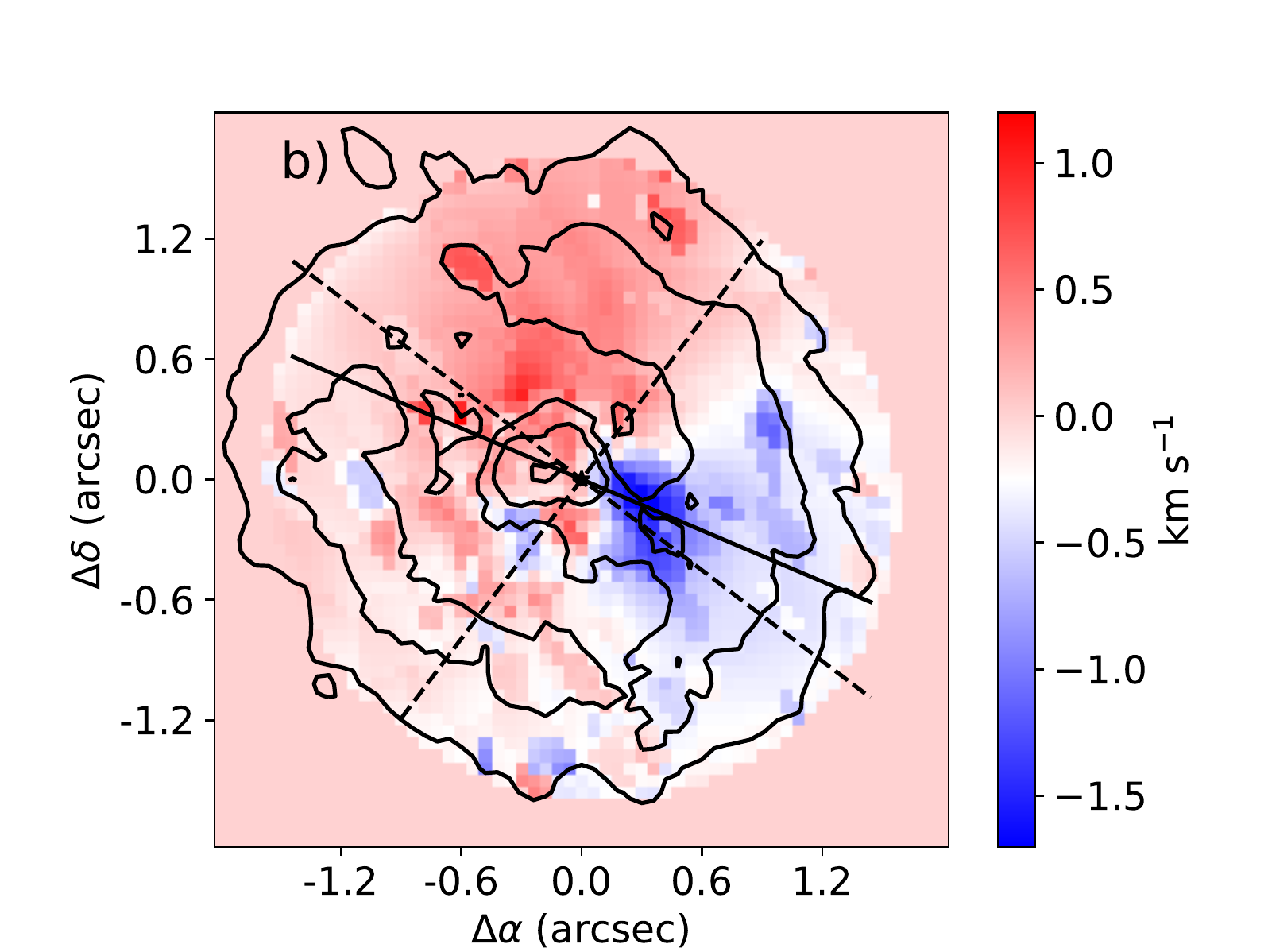}\includegraphics[scale=0.37, trim=5mm 0mm 15mm 10mm, clip]{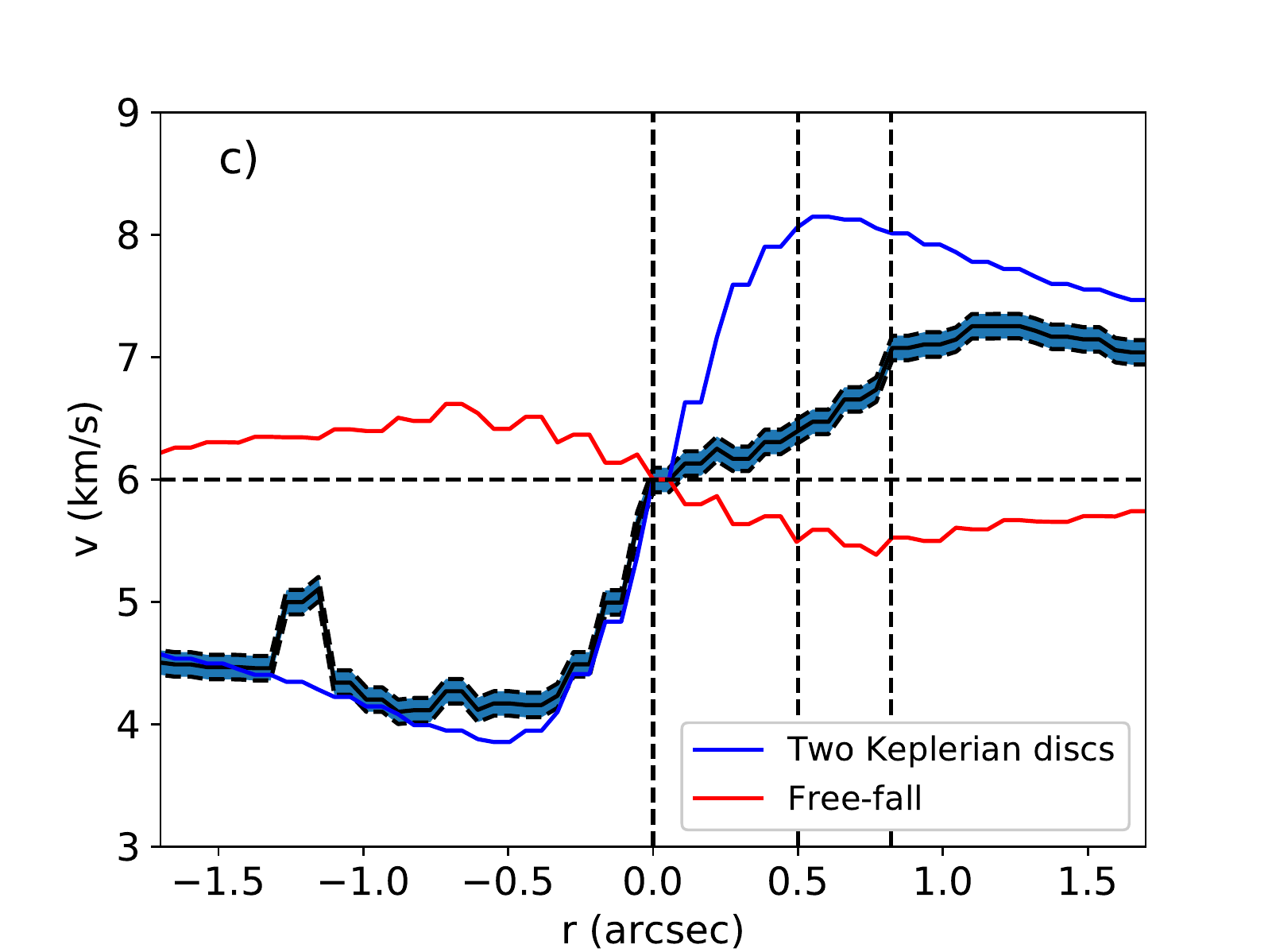} \\
\caption{(a) HCO$^+$ first moment map. The two dashed lines show the direction of the disk major and minor axes. The arrow marks the position of the radio jet \citep{Rodriguez2014}. The blue circle marks the position of a velocity anomaly associated with the bridge. (b) Residuals of a model consisting of two Keplerian disks (see Section \ref{Sec:Gas_kin}). Solid contours depict 5$\sigma$ to 25$\sigma$ levels of the integrated line intensity. The black solid line shows the direction of the cut through the bridge used for panel (c). (c) Cut along the velocity anomaly  for the HCO$^+$ first moment map (black lines with dashed region), a two Keplerian disks model as in panel (b) (blue line), and a free-falling component (red line). The dashed vertical lines mark the positions of the center, the bridge, and the inner edge of the dust disk.}
\label{Fig:v_field}
\end{figure*}

We show in Figure \ref{Fig:v_field}a the first moment map of HCO$^+$ emission. The outer parts of the disk resemble those of a Keplerian disk. However, twisted isophotes are observed toward the center ($r <0\arcsec$.6). The aforementioned twisting can be reproduced by two misaligned Keplerian disks as proposed by \cite{Hashimoto2011}. A radial outflow would produce a similar effect \citep[][]{Rosenfeld2014}. We adopt the misaligned disks as the most plausible explanation, as it is consistent with NIR observations by \cite{Hashimoto2011}. In Figure \ref{Fig:v_field}b, we show the residual map that results from subtracting a two Keplerian disks model from the HCO$^+$  first moment map. The model consists of an outer disk extending from 350 to 45 au, with $i = 26^{\circ}$ and PA = 37$^{\circ}$, and an inner one extending from 45 au inward with $i = 43 ^{\circ}$ and PA = 65$^{\circ}$ \citep{Hashimoto2011}, around a 2.4 $M_{\sun}$ star, and the final image is convolved with  the observational beam. This simplistic model provides a reasonable fit across the disk, except in the southwestern region, where strong residuals are observed. Visual inspection of Figure \ref{Fig:v_field}a, shows that there is an anomaly (blue circle) in the velocity field that is spatially coincident with the bridge (BR, Figure \ref{Fig:v_field}b). This anomaly is due to high-velocity wings in the line profile of HCO$^+$ at the bridge position (see Figure \ref{Fig:SpectralComparison}b). The red-shifted peak is compatible with Keplerian rotation, while the blue-shifted one requires a different mechanism, most likely infalling gas. To test the possibility that the blue shifted wing is tracing infalling material, we built a pure free-falling velocity field following equation

\begin{equation}
v_{free-fall} = \sqrt{2GM_{*}\times \left(\frac{1}{r}-\frac{1}{r_{start}}\right)}
\end{equation}
where $r_{\rm start}$ is the starting position for free-fall accretion, and we adopted $r_{\rm start}$ = 120 au (accretion starts at the inner edge of the dust ring,  $ r \sim$ 0$\arcsec$.7). We assumed that infall stops at 40 au. Cuts in the first-moment map, and in our convolved models are shown in Figure \ref{Fig:v_field}c. The free-falling model has a velocity of 5.5 km $\rm s^{-1}$ at the bridge position, compared to the 5.3 km $\rm s^{-1}$ derived from the Gaussian fit. Therefore, some of the material in the bridge seems to be infalling at a velocity close to free-fall. Smaller velocities, such as the speed of sound, which is typically 0.05-0.1 times the Keplerian velocity, cannot account for the blue-shifted component. Assuming that  this component is due to accretion, we derive a mass accretion 3$\rm \times 10^{-8}$ to 3$\rm \times 10^{-7}$ $M_{\odot}$ $\rm yr^{-1}$ (for $X$(HCO$^+$) = 10$\rm ^{-9}$ to 10$\rm ^{-8}$), comparable to a value of 1.2$\rm \times 10^{-7}$ $M_{\odot}$ $\rm yr^{-1}$ by \cite{Salyk2013}. 

Radial inflows through dust cavities have been invoked before to explain different features observed in a few protoplanetary disks, such as gas and dust streamers and twisted isophotes \citep{Cassasus2013, Dutrey2014, Perez2015, Zhang2015, Loomis2017, Mendigutia2017, Walsh2017}. In particular, \cite{Zhang2015} proposed radial inflows to explain absorption features observed in CO ro-vibrational lines in AA Tau, with material moving at velocities close to free-fall in regions between the inner and outer disk. The presence of gap-crossing radial inflows could explain the high accretion levels derived for some transitional disks, such as those observed in AB Aur  \citep{GarciaLopez2006, Salyk2013}. A compact and collimated bipolar jet was detected by \citet{Rodriguez2014} in the continuum at 7 mm suggesting the existence of a disk-jet system. In order to sustain it, gas must flow through the cavity and reach the inner regions of the system, otherwise accretion would drain the inner disk on very short timescales. 

We cannot discard the possibility that the non-Keplerian component observed along the bridge is due to gas out of the disk plane, in which case, depending on the exact location of the bridge, it might be interpreted as an inflow or an outflowing stream.

Previous work by this team proposed that the asymmetric dust ring is caused by dust trapping in a decaying gas vortex formed at the outer edge of the gap carved by a massive planet, coincident with the position of the bridge \citep{Fuente2017}. Higher-spatial resolution observations, with facilities such as ALMA, are needed to understand the connection between the HCO+ bridge in the cavity and the putative planet responsible for the outer dust ring. 

\section{Summary and Conclusions}
We used the NOEMA interferometer to map with high angular resolution the HCO$\rm ^+$ and HCN J = 3-2 lines. Based on these interferometric data, we explored the chemistry and dynamics of the molecular gas component in the AB Aur disk, and its relationship with the decaying gas vortex found in it. We detected ring-like emission in HCN, spatially coincident with dust emission. HCO$^+$ shows a ring-like structure in the outer regions, plus a compact source at the center. We suggest that the HCO$^+$/HCN ratio is a useful tool to discern the presence of dense gas in dust cavities.

Our most prominent result is the detection of a filamentary HCO$^+$ emission structure that connects the outer disk with the inner regions. A simple model consisting of two misaligned Keplerian disks with an infall component along the bridge can reproduce our observations, suggesting the presence of active accretion within the dust cavity. Yet, alternative explanations involving gas out of the disk plane cannot be ruled out.  The possible connection of the observed bridge with a planet formation scenario requires further research.

\acknowledgements
We thank the anonymous referee for a fruitful report that helped us to improve the discussion of our results. 
We thank the Spanish MINECO for funding support from
AYA2016-75066-C2-1/2-P, AYA2012-32032 and ERC under ERC-2013-SyG, G. A. 610256 NANOCOSMOS.
This work was supported by the Programme National $``$Physique et Chimie du Milieu Interstellaire$\arcsec$
(PCMI) of CNRS/INSU with INC/INP co-funded by CEA and CNES. S.P.T.M. acknowledges to the European Union's Horizon 2020 research and innovation program for funding support given under grant agreement No.639459 (PROMISE).

\end{document}